\documentclass[aps,prl,superscriptaddress,twocolumn,twoside,a4paper,floatfix]{revtex4}
\usepackage{hyperref,graphicx,amsmath,amssymb,mathpazo,color}
\usepackage[normalem]{ulem}

\usepackage{amsfonts,epsfig}
\usepackage{latexsym}
\usepackage{epsfig}
\usepackage{amsmath}
\usepackage{amssymb}
\usepackage{amsfonts}
\usepackage{amsthm}
\usepackage{mathrsfs}
\usepackage{natbib}
\usepackage{color,verbatim,graphics}
\usepackage{psfrag}
\DeclareMathAlphabet{\mathrsfs}{U}{rsfs}{m}{n}
\DeclareMathAlphabet{\mathpzc}{OT1}{pzc}{m}{it}
\DeclareMathAlphabet{\matheus}{U}{eus}{m}{n}
\DeclareMathAlphabet{\mathbbold}{U}{bbold}{m}{n}

\newcommand{\ba}{\begin{eqnarray}}
\newcommand{\be}{\begin{equation}}
\newcommand{\ee}{\end{equation}}

\newcommand{\ea}{\end{eqnarray}}
\newcommand{\ban}{\begin{eqnarray*}}
\newcommand{\ean}{\end{eqnarray*}}
\newcommand{\Tr}{\operatorname{Tr}}

\newcommand{\ket}[1]{|#1\rangle}
\newcommand{\bra}[1]{\langle#1|}

\newcommand{\expect}[1]{\langle#1\rangle}

\newcommand{\ie}{{\it{i.e.} }}

\newcommand*\diff{\mathop{}\!\mathrm{d}}

\begin{document}

\title{Genuine hidden quantum nonlocality}

\author{Flavien Hirsch}
\affiliation{D\'epartement de Physique Th\'eorique, Universit\'e de Gen\`eve, 1211 Gen\`eve, Switzerland}
\author{Marco T\'ulio Quintino}
\affiliation{D\'epartement de Physique Th\'eorique, Universit\'e de Gen\`eve, 1211 Gen\`eve, Switzerland}
\author{Joseph Bowles}
\affiliation{D\'epartement de Physique Th\'eorique, Universit\'e de Gen\`eve, 1211 Gen\`eve, Switzerland}
\author{Nicolas Brunner}
\affiliation{D\'epartement de Physique Th\'eorique, Universit\'e de Gen\`eve, 1211 Gen\`eve, Switzerland}
\affiliation{H.H. Wills Physics Laboratory, University of Bristol, Bristol, BS8 1TL, United Kingdom}

\begin{abstract}
The nonlocality of certain quantum states can be revealed by using local filters before performing a standard Bell test. This phenomenon, known as hidden nonlocality, has been so far demonstrated only for a restricted class of measurements, namely projective measurements. Here we prove the existence of genuine hidden nonlocality. Specifically, we present a class of two-qubit entangled states, for which we construct a local model for the most general local measurements (POVMs), and show that the states violate a Bell inequality after local filtering. Hence there exist entangled states, the nonlocality of which can be revealed only by using a sequence of measurements. Finally, we show that genuine hidden nonlocality can be maximal. There exist entangled states for which a sequence of measurements can lead to maximal violation of a Bell inequality, while the statistics of non-sequential measurements is always local.
\end{abstract}

\maketitle

Performing local measurements on separated entangled particles can lead to nonlocal correlations, as witnessed by the violation of a Bell inequality \cite{bell}. 
This phenomenon, termed quantum nonlocality, has received strong experimental confirmation. Moreover, entanglement and nonlocality are now viewed as fundamental aspects of quantum theory, and play a prominent role in quantum information \cite{cc, review}.

However, 50 years after the discovery of Bell's theorem, we still do not fully understand the relation between entanglement and nonlocality, although significant progress was made \cite{review}. In particular, the most natural question of which entangled states can lead to nonlocal correlations and which ones cannot, is still open. While it is known that nonlocality is a generic feature for pure entangled states \cite{gisin,pr}, the situation for mixed states turns out to be much more complex. 
First, Werner \cite{werner} showed that there exist mixed entangled states (so called Werner states) that admit a local model for projective measurements. However, it could still be the case that such states violate a Bell inequality when more general measurements (POVMs) are considered. Motivated by this question, Barrett \cite{barrett} showed that certain noisy Werner states (but nevertheless entangled) admit a local model even when POVMs are considered (see also \cite{maf}). 

Another twist to this question was given in Ref. \cite{popescu,gisinfilter}, proposing Bell tests where observers perform a sequence of measurements---rather than a single measurement. Notably, Popescu \cite{popescu} showed that Werner states of local dimension $d \geq 5$ can violate a Bell inequality when judicious local filters are applied to the state before performing a standard Bell test. Hence, the local filters reveal the hidden nonlocality of the quantum state. Importantly, the use of local filters does not open any loophole, since the choice of local measurement settings (for the second measurement) can be performed after applying the filters \cite{popescu,zukowski,teufel}. While this result shows that sequential measurements can be beneficial in Bell tests, it raises the question of whether they are necessary. Indeed, the crucial point here is that hidden nonlocality has been so far demonstrated only for a restricted class of measurements, namely projective measurements. Specifically the Werner states considered by Popescu admit a local model for projective measurements, but could in principle violate a Bell inequality when POVMs are considered. Indeed, POVMs are proven to be relevant in Bell tests, as they can increase Bell violation compared to projective measurements \cite{tamas}. Hence, this raises the question of whether there exists \emph{genuine hidden nonlocality}. That is, do there exist  entangled states, the nonlocality of which can be observed only if sequential measurements are used?

Here we prove the existence of genuine hidden nonlocality. Specifically, we start by presenting a simple class of two-qubit entangled states, for which we construct a local model for POVMs, \ie arbitrary non-sequential measurements. Next, we show that these states violate the Clauser-Horne-Shimony-Holt (CHSH) \cite{chsh} Bell inequality when a judiciously chosen sequence of measurements is performed. Hence, this shows that sequential measurements outperform non-sequential ones, and that the nonlocality of certain entangled states can be revealed only through a sequence of measurements. Moreover, our construction provides the simplest example of hidden nonlocality known so far. A central tool for deriving our result is a technique which allows us, starting from a local model for simulating dichotomic projective measurements on a given state, to construct a local model for simulating POVMs on a related (but in general different) state. Finally, we demonstrate that genuine hidden nonlocality can be maximal. Specifically, we present a simple class of qutrit-qutrit entangled states which admit a local model for POVMs, but violate maximally the CHSH inequality when a sequence of measurements is used. Hence, such states are useful resources for information-theoretic tasks based on nonlocality \cite{cc, review}, although they seem useless at first sight.
These results highlight novel aspects of the subtle relation between entanglement and nonlocality.

We start by introducing the scenario and notations. Consider a bipartite Bell scenario in which distant parties, Alice and Bob, perform local measurements on an entangled state $\rho$ of local Hilbert space dimension $d$. The choice of measurement setting is denoted by $x$ for Alice ($y$ for Bob), and the measurement outcome by $a$ ($b$ for Bob). Each setting is represented by a collection of positive operators acting on $\mathbb{C}^d$ denoted here $M_{a|x}$ and $M_{b|y}$ satisfying the relations $\sum_a M_{a|x}= \openone$ and $\sum_b M_{b|y}= \openone$, where $\openone$ denotes the identity operator in dimension $d$. The experiment is then characterized by the joint probability distribution 
\ba p(ab|xy) = \Tr(M_{a|x}  \otimes M_{b|y} \, \rho). \ea
If the distribution $p(ab|xy)$ violates (at least) one Bell inequality, the state $\rho$ is said to be nonlocal. If on the other hand the distribution admits a decomposition 
\ba\label{loc} p(ab|xy) = \int \diff \lambda \omega(\lambda) p(a|x\lambda) p(b|y\lambda) \ea
for all possible measurements, the state $\rho$ admits a local model, and cannot violate any Bell inequality. Here $\lambda$ represents the local hidden variable, distributed according to the density $\omega(\lambda)$. We will consider two separate cases. First, when a decomposition of the form \eqref{loc} can be found for all projective measurements (\ie  $M_{a|x}^2 = M_{a|x}$ and $M_{b|y}^2 = M_{b|y}$) we say that $\rho$ is local for projective measurements. Second, if a decomposition of the form \eqref{loc} can be found for all POVMs (arbitrary non-sequential measurements), we say that $\rho$ is local for POVMs.

So far, we have considered a Bell scenario in which each party performs a single measurement on its particle. One can however consider a more general measurement scenario, in which each party performs a sequence of measurements \cite{popescu,gisinfilter}. For instance, upon receiving their particle, the parties apply a local filtering. In the case that the filtering succeeds on both sides, the parties now hold the 'filtered' state
\ba  \tilde{\rho} = \frac{1} {N} [ (F_A \otimes F_B) \rho ( F_A^{\dagger} \otimes F_B^{\dagger})] \ea
where $N= \Tr[ (F_A \otimes F_B) \rho ( F_A^{\dagger} \otimes F_B^{\dagger})] $ is a normalization factor, and $F_A$ and $F_B$ are positive operators acting on $\mathbb{C}^d$ representing the local filtering of Alice and Bob. Finally, the parties perform local measurements on $\tilde{\rho} $ and can test a Bell inequality. 
Here we will see that such a sequence of measurements is necessary in certain cases. More precisely, there exist entangled quantum states, the nonlocality of which can only be revealed by performing sequential measurements. Thus such states exhibit genuine hidden nonlocality.

To demonstrate our main result, we proceed in several steps. First, we consider a simple class of entangled two-qubit states, of the form
\ba \label{s1}  \rho = q \Psi_- + (1-q) \ket{0}\bra{0}  \otimes \frac{\openone}{2} \ea
where $\Psi_- = \ket{\psi_-}\bra{\psi_-}$ denotes the projector on the singlet state $\ket{\psi_-}=(\ket{0,1}-\ket{1,0})/\sqrt{2}$, and $0\leq q \leq 1$. 
Building upon the models discussed in Refs \cite{gisingisin,degorre}, we will see now that state \eqref{s1} admits a local model for projective measurements when $q\leq 1/2$, although it is entangled for all $q>0$. Specifically, Alice and Bob receive as input a vector $\vec{x}$ and $\vec{y}$, and should simulate the statistics of measuring qubit observables $\vec{x}\cdot \vec{\sigma}$ and $\vec{y}\cdot \vec{\sigma}$ on $\rho$; here $ \vec{\sigma}$ denotes the vector of Pauli matrices, hence the measurement outcomes are $\pm 1$.

{\bf Protocol 1.} Alice and Bob share a 3-dimensional unit vector $\vec{\lambda}$, uniformly distributed on the sphere. Upon receiving $\vec{x}$, Alice tests the shared vector $\vec{\lambda}$. With probability $| \vec{x}\cdot \vec{\lambda} |$, she 'accepts' $\vec{\lambda}$, and outputs $a= -\text{sign}(\vec{x}\cdot \vec{\lambda})$; otherwise, she outputs $a=\pm1$ with probability $(1 \pm \bra{0}\vec{x}\cdot \vec{\sigma} \ket{0})/2$. Bob simply outputs $b=\text{sign}(\vec{y}\cdot \vec{\lambda})$. 

The protocol consists of two parts. First, when Alice accepts $\vec{\lambda}$, which occurs on average with probability $1/2$ (independently of $\vec{x}$), $\vec{\lambda}$ is distributed according to the density $\omega(\vec{\lambda})= | \vec{x}\cdot \vec{\lambda} |/ 2\pi$ \cite{gisingisin,degorre}. In this case, the correlation between Alice's and Bob's outcomes is
\ba \expect{ab} &=&
- \frac{1}{2\pi} \int \diff \vec{\lambda} | \vec{x}\cdot \vec{\lambda} | \text{sign}(\vec{x}\cdot \vec{\lambda}) \text{sign}(\vec{y}\cdot \vec{\lambda}) \nonumber \\
&=& -  \vec{x}\cdot \vec{y} \ea
where the integral is taken over the sphere. As the marginals are uniform, \ie $\expect{a}=\expect{b}=0$, we recover the singlet correlations. Second, when Alice rejects $\vec{\lambda}$, she simulates locally the statistics of the state $\ket{0}$, while Bob's outcome is uncorrelated. Hence the model reproduces exactly the statistics of the state \eqref{s1} for $q =1/2$, \ie $\expect{ab}=(-  \vec{x}\cdot \vec{y})/2$, $\expect{a}=x_z /2$ and $\expect{b}=0$. The case $q <1/2$ is a trivial extension.

At this point, it is relevant to note that after local filtering, the state \eqref{s1} violates the CHSH inequality $|S|\leq 2$ \cite{chsh}, where $S = E_{1,1}+E_{1,2}+E_{2,1}-E_{2,2} $ and $E_{x,y}= \sum_{a,b= \pm 1} (ab) p(ab|xy)$. Specifically, applying filters of the form 
\ba \label{filters} F_{A} = \epsilon \ket{0} \bra{0} + \ket{1} \bra{1} \, , \,\, F_{B} = \delta \ket{0} \bra{0} + \ket{1} \bra{1} \ea
with $\delta = \epsilon / \sqrt{q}$ to state \eqref{s1}, we obtain the filtered state 
\ba \tilde{\rho} \simeq \sqrt{q} \, \Psi_- + (1-\sqrt{q}) \frac{\ket{0,1}\bra{0,1}+ \ket{1,0}\bra{1,0}}{2} + \mathcal{O}(\epsilon^2) \nonumber \ea  
which violates CHSH up to $S=2 \sqrt{1+q}$ (for $\epsilon \rightarrow 0$) according the Horodecki criterion \cite{horodecki}. Note that filters \eqref{filters} are optimal here \cite{verstraete}. Hence the state \eqref{s1} exhibits hidden nonlocality for projective measurements. This shows that hidden nonlocality exists for two-qubit states---the previous example \cite{popescu} considered Werner states of local dimension $d \geq 5 $. However, at this point we cannot ensure that the state \eqref{s1} is local for all non-sequential measurements, since Bell violation could in principle be obtained using POVMs. Nevertheless, we will now build upon the above construction to present a state featuring genuine hidden nonlocality.

Our main tool is a protocol for constructing a state which admits a local model for POVMs. Specifically, starting from a state $\rho_0$ of local dimension $d$ which is local for dichotomic projective measurements, we construct the state
\ba\label{rhoPOVM} \rho' &=& \frac{1}{d^2} [\rho_0 + (d-1)(\rho_A \otimes  \sigma_B +  \sigma_A \otimes \rho_B ) \nonumber \\
& & + (d-1)^2 \sigma_A \otimes  \sigma_B] \ea
which is local for POVMs. Here $\sigma_{A,B}$ are arbitrary $d$-dimensional states, and $\rho_{A,B}= \Tr_{B,A}(\rho_0)$. 

Alice receives as input a POVM $\{M_a \} $ (from now on we omit the subscript $x$). Without loss of generality, each POVM element $M_a$ can be taken to be proportional to a rank-one projector $P_a$ (see e.g. \cite{barrett}), \ie $M_a = \alpha_a P_a$ with $\alpha_a \geq 0$ and $\sum_a \alpha_a = d$ by normalization of the POVM. Bob receives POVM $\{M_b \} $ (with $M_b = \beta_b P_b$). The protocol is explained below for Alice; Bob follows the same procedure.

{\bf Protocol 2.} (i) Alice chooses projector $P_a$ with probability $\alpha_a / d$ (note that $\sum_a \alpha_a/d = 1$). (ii) She   
simulates the dichotomic projective measurement $\{P_a, \openone -P_a\}$ on state $\rho$. (iii) If the output of the simulation corresponds to $P_a$, she outputs $a$. (iv) Otherwise, she outputs (any) $a$ with probability $\Tr(M_a \sigma_A)$. 

Let us now show that the protocol simulates $\rho'$. Note first that the probability that Alice outputs in state (iii) is given by $\sum_a \alpha_a/d \Tr(P_a \rho_A)=1/d$. We will now evaluate the probability that the parties output given values $a$ and $b$ in the protocol. Four cases are possible: 1. Both Alice and Bob output in step (iii), which occurs with probability $ \frac{\alpha_a}{d} \frac{\beta_b}{d} \Tr(P_a \otimes P_b \rho_0) = \frac{1}{d^2}\Tr(M_a \otimes M_b \rho_0) $; 2. Alice outputs in step (iii) and Bob in step (iv), which occurs with probability 
\ba \sum_k \frac{\alpha_a}{d} \frac{\beta_k}{d} \Tr(P_a( \openone - P_k) \rho_0) \Tr(M_b \sigma_B) \nonumber \\ 
=  \frac{d-1}{d^2} \Tr(M_a \rho_A) \Tr(M_b \sigma_B)  \ea
3. Alice outputs in step (iv) and Bob in step (iii) has probability $\frac{d-1}{d^2} \Tr(M_a \sigma_A) \Tr(M_b \rho_B) $; 4. Both Alice and Bob output in step (iv), which occurs with probability $\frac{(d-1)^2}{d^2} \Tr(M_a \sigma_A) \Tr(M_b \sigma_B) $. Altogether, we have that $ p(ab) = \Tr(M_a \otimes M_b \rho') $. Hence the model reproduces the statistics of arbitrary POVMs on the state $\rho'$.


We are now ready to show our main result. We use protocol 2 with $\rho_0$ given by the state of Eq. \eqref{s1}, which is local for projective measurements for $q\leq 1/2$, and choosing $\sigma_{A,B}= \ket{0}\bra{0} $, we obtain a state of the form
\ba \label{s11} \rho_G &=& \frac{1}{4} [ q  \Psi_- +  (2-q) \ket{0}\bra{0} \otimes  \frac{\openone}{2} + q \frac{\openone}{2}  \otimes \ket{0}\bra{0} \nonumber \\
& & + (2-q) \ket{0,0}\bra{0,0}]  \ea
which is local for POVMs by construction for $q \leq 1/2$. Nevetheless $\rho_G$ is nonlocal for any $q>0$ when an appropriate sequence of measurements is used. In particular, applying filters of the form \eqref{filters} with $\delta = \epsilon / \sqrt{q} $ to state $\rho_G$, we obtain
\ba \tilde{\rho}_G \simeq \frac{\sqrt{q}}{2} \, \Psi_- + (1-\frac{\sqrt{q}}{2}) \frac{\ket{0,1}\bra{0,1}+ \ket{1,0}\bra{1,0}}{2} + \mathcal{O}(\epsilon^2) \nonumber \ea  
which violates CHSH up to $S=2 \sqrt{1 + q/4}$ (for $\epsilon \rightarrow 0$) according the criteria of Ref. \cite{horodecki}. Hence, sequential measurements are necessary to reveal the nonlocality of the state \eqref{s11}, which therefore exhibits genuine hidden nonlocality.

Finally, we present a stronger version of this phenomenon, showing that there exist quantum states with genuine and maximal hidden nonlocality. That is, although the state admits a local model for POVMs, it violates maximally the CHSH inequality when sequential measurements are used, as the state after filtering is a pure singlet state. 

We start here by considering the qutrit-qubit state
\ba \label{erasure}  \rho_E = q \Psi_- +  (1-q) \ket{2}\bra{2}  \otimes \frac{\openone_2}{2} \ea
where $\openone_2$ denotes the identity in the $\ket{0},\ket{1} $ qubit subspace. This state is usually referred to as the 'erasure state', as it can be obtained by sending half of a singlet state $ \Psi_-$ through an erasure channel; with probability $q$ the singlet state remains intact, and with probability $(1-q)$ Alice's qubit  is lost and replaced by the state $\ket{2}\bra{2}$ (orthogonal to the qubit subspace).

The state \eqref{erasure} is local for dichotomic projective measurements when $q \leq 1/2$. Consider Alice receiving an observable with eigenvalues $\pm1$, which can always be written as an operator of the form $c_0 \vec{x} \cdot \vec{\sigma} + c_1 \openone_2 +  R$, where $c_0,c_1 \in [0,1]$, operators $\vec{x} \cdot \vec{\sigma}$  and $\openone_2$ act on the $ \ket{0},\ket{1} $ qubit subspace, 
and operator $R$ has no support in the qubit subspace. The protocol is similar to protocol 1. Alice and Bob share a vector $\vec{\lambda}$. Alice accepts $\vec{\lambda}$ with probability $| \vec{x}\cdot \vec{\lambda} |$, in which case she outputs $a= -\text{sign}(\vec{x}\cdot \vec{\lambda})$ with probability $c_0$, and a random bit otherwise. If she rejects $\vec{\lambda}$, she outputs $\pm1$ with probability $[1 \pm (c_1 + \Tr{R} )/2]/2$. Bob receives observable $\vec{y} \cdot \vec{\sigma}$ and outputs $b=\text{sign}(\vec{y}\cdot \vec{\lambda})$. 

Noting that Alice accepts $\vec{\lambda}$ with probability 1/2 on average, we obtain $\expect{ab} = -c_0 ( \vec{x} \cdot \vec{y})/2 $, $\expect{a}= (c_1 + \Tr{R})/2$, and $\expect{b}=0$, which is the statistics of dichotomic projective measurements on state $\rho_E$ for $q=1/2$. Next, we apply protocol 2 to $\rho_E$, taking $\sigma_{A,B}= \ket{2}\bra{2} $. Hence the state 
\ba \label{s2} \rho_{GM} &=& \frac{1}{9} [ q \Psi_- +  (3-q) \ket{2}\bra{2} \otimes \frac{\openone_2}{2} + 2q \frac{\openone_2}{2}  \otimes \ket{2}\bra{2} \nonumber \\ & & + (6-2q) \ket{2,2}\bra{2,2}]  \ea
is local for POVMs for $q \leq 1/2$. To reveal the nonlocality of the above state, we apply filters of the form $F_A = F_B = \ket{0}\bra{0} + \ket{1}\bra{1}$. Hence after successful filtering, we obtain a pure singlet state, \ie $\tilde{\rho}_{GM} = \Psi_-$. By performing suitable measurements on $\tilde{\rho}_{GM}$, Alice and Bob can now get maximal violation of the CHSH inequality, \ie   $S=2 \sqrt{2}$ \cite{tsirelson}. Therefore, the state \eqref{s2} has genuine and maximal hidden nonlocality. 

Note also that applying the above filters to the erasure state \eqref{erasure} gives a pure singlet state for any $q>0$. Thus the erasure state with $0<q\leq 1/2$ has hidden nonlocality for dichotomic measurements. Moreover, for $q\leq 1/6$, the erasure state admits a local model for projective measurements, as can be shown by using protocol 2 \footnote{Apply protocol 2 to $\rho_E$ with $q=1/2$ and take $\sigma_A= \ket{2}\bra{2}$. Note that the protocol must be applied solely on Alice's side. Since Bob holds a qubit, the simulation model for dichotomic measurements includes already all projective measurements.}. Hence, such states feature hidden nonlocality for projective measurements.

To summarize, we have shown the existence of genuine hidden nonlocality. That is, there exist entangled quantum states the nonlocality of which can be revealed only via sequential measurements. In certain cases, this nonlocality can even be maximal.


In the present paper, we have focused on Bell tests in which a single copy of an entangled state is measured in each run of the experiment. It is however also relevant to consider the case in which several copies of the state can be measured jointly in each run \cite{peres,liang,masanes,liang2}. Notably, it has been shown recently that nonlocality can be super-activated in this scenario \cite{carlos}. That is, by performing judicious joint measurements on sufficiently many copies of a state $\rho$, it becomes possible to violate a Bell inequality (with non-sequential measurements) although the state $\rho$ admits a local model for POVMs. More generally, this phenomenon occurs for any entangled state $\rho $ that is useful for teleportation \cite{cavalcanti13}. It is thus interesting to ask whether the nonlocality of the states considered here could also be revealed by allowing for many copies to be measured jointly. However, the current results on super activation of quantum nonlocality do not detect the states presented here \footnote{Currently, the best result shows that super-activation is possible when the state has singlet fidelity greater than $1/d$ \cite{cavalcanti13}, which is not the case for states \eqref{s11} and \eqref{s2} (when $q \leq 1/2$), and state \eqref{s1} (for $q \leq 1/3$).}, thus leaving the question open. Another point worth mentioning is activation of nonlocality in quantum networks. 
It would also be relevant to see whether the nonlocality of the states presented here can be activated by placing several copies of them in a quantum network \cite{dani}. Concerning the erasure state, Ref. \cite{dani2} shows that it is a nonlocal resource when placed in a tripartite network, hence the local model constructed here confirms that activation of nonlocality does indeed occur.

Finally, an interesting open question is whether there exist entangled states for which nonlocality cannot be observed even considering sequential measurements on an arbitrary number of copies of the state.

\emph{Acknowledgements.} We thank Y.-C. Liang for discussions. We also thank A. Acin, F. Brand\~ao, D. Cavalcanti, N. Gisin, T. Maciel. We acknowledge financial support from the Swiss National Science Foundation (grant PP00P2\_138917) and the EU DIQIP.

\end{document}